\documentclass[%
 aip,
 amsmath,amssymb,
 reprint,%
]{revtex4-1}

\usepackage{siunitx}
\usepackage{mathtools}
\usepackage[normalem]{ulem}
\usepackage{color}

\begin{document}

\title{On explosive boiling of a multicomponent Leidenfrost drop}

\author{Sijia Lyu}
\affiliation{Center for Combustion Energy, Key Laboratory for Thermal Science and Power Engineering of Ministry of Education, Department of Energy and Power Engineering, Tsinghua University, 100084 Beijing, China}
\author{Huanshu Tan} 
\affiliation{Department of Chemical Engineering, University of California, Santa Barbara, USA}
\author{Yuki Wakata}
\affiliation{Center for Combustion Energy, Key Laboratory for Thermal Science and Power Engineering of Ministry of Education, Department of Energy and Power Engineering, Tsinghua University, 100084 Beijing, China}
\author{Xianjun Yang}
\affiliation{Center for Combustion Energy, Key Laboratory for Thermal Science and Power Engineering of Ministry of Education, Department of Energy and Power Engineering, Tsinghua University, 100084 Beijing, China}
\author{Chung K. Law}
\affiliation{Department of Mechanical and Aerospace Engineering, Princeton University, Princeton, NJ 08544, USA}
\author{Detlef Lohse}
\affiliation{Physics of Fluids Group, MESA$^{+}$ Institute  and J.M. Burgers Centre for Fluid Dynamics, University of Twente, P.O. Box 217, 7500AE Enschede, The Netherlands}
\affiliation{Max Planck Institute for Dynamics and Self-Organization, 37077 G\"ottingen, Germany}
\author{Chao Sun}
\email{chaosun@tsinghua.edu.cn}
\affiliation{Center for Combustion Energy, Key Laboratory for Thermal Science and Power Engineering of Ministry of Education, Department of Energy and Power Engineering, Tsinghua University, 100084 Beijing, China}
\affiliation{Physics of Fluids Group, MESA$^{+}$ Institute  and J.M. Burgers Centre for Fluid Dynamics, University of Twente, P.O. Box 217, 7500AE Enschede, The Netherlands}
\affiliation{Department of Engineering Mechanics, School of Aerospace Engineering, Tsinghua University, Beijing 100084, China}

\begin{abstract}
The gasification of multicomponent fuel drops is relevant in various energy-related technologies. An interesting phenomenon associated with this process is the self-induced explosion of the drop, producing a multitude of smaller secondary droplets, which promotes overall fuel atomization and, consequently, improves the combustion efficiency and reduces emissions of liquid-fueled engines. Here, we study a unique explosive gasification process of a tri-component droplet consisting of water, ethanol, and oil (``ouzo"), by high-speed monitoring of the entire gasification event taking place in the well-controlled, levitated Leidenfrost state over a superheated plate. It is observed that the preferential evaporation of the most volatile component, ethanol, triggers nucleation of the oil microdroplets/nanodroplets in the remaining drop, which, consequently, becomes an opaque oil-in-water microemulsion. The tiny oil droplets subsequently coalesce into a large one, which, in turn, wraps around the remnant water. Because of the encapsulating oil layer, the droplet can no longer produce enough vapor for its levitation, and, thus, falls and contacts the superheated surface. The direct thermal contact leads to vapor bubble formation inside the drop and consequently drop explosion in the final stage. 
\end{abstract}


\maketitle

Drop gasification, possibly followed by combustion, is a major component in daily practices and technologies such as vehicle propulsion via internal combustion engines, spray cooling and painting, and various energy conversion industries~\cite{demirbas2005potential,law2010combustion,Detlef_physio}. 
These drops are frequently multicomponent, governed by complex gasification and combustion processes, due to different volatilities and mutual solubilities of the liquid components~\cite{christy2011flow, diddens2017evaporating}, the interaction between the substrate and the drop~\cite{bourrianne2019cold, tran2013droplet, nair2014leidenfrost, Duan2014, gao2018evaporation}, and different boiling regimes at varying local temperatures~\cite{shirota2016dynamic, tran2012drop, staat2015phase}.

Recent studies on miscible and emulsified fuel drops~\cite{tsue1998effect,takashima2005evaporation,calabria2007combustion,watanabe2010experimental,mura2012study}  have shown that, for certain optimal compositions, the drop can undergo self-induced (micro)explosion upon gasification, producing multitudes of much smaller, secondary drops which promote the overall fuel atomization, and, consequently {improve the combustion efficiency and reduce emissions}~\cite{law1982recent,sirignano1983fuel,kadota2007microexplosion}. It has been further shown that the explosion is especially intense for emulsion drops, such as water dispersed in a heavy oil, whose boiling temperature is substantially higher than the limit of superheat of water. However, its practical utilization is frequently constrained by the requirement of substantial amount of surfactants to achieve phase stability for storage.

Capitalizing on the concept of inducing the superheating of one volatile liquid component by another high-boiling-point immiscible component, as observed in the above water-in-oil emulsion, the present study is motivated by the desire to explore alternate fuel mixture systems exhibiting microexplosion capabilities. In particular, in one of these smart systems, the mixture can phase separate during the course of drop gasification, and subsequently self-explode through the superheating of the volatile liquid components by the phase-separated, relatively less volatile, high-boiling point component. 
The multicomponent liquid we selected is the ouzo mixture, which consists of ethanol, water, and a high boiling point trans-anethole oil, and has been extensively studied in physicochemical hydrodynamic problems ~\cite{tan2016evaporation, lu2017universal, moerman2018emulsion,otero2018compositional,li2019bouncing,Detlef_physio}.
The ternary phase diagram of the solution provides information on phase separation (SI Appendix, Fig. S1). In particular, it quantifies how increasing the water-ethanol ratio can lead to oil droplet nucleation because of the progressive reduction of oil solubility.
Here, we use the Leidenfrost arrangement~\cite{quere2013leidenfrost} by placing a small ouzo drop on a superheated substrate such that the drop levitates on its own vapor layer~\cite{leidenfrost1756aquae, quere2013leidenfrost} and is heated through it. We shall investigate the entire boiling process of such an ouzo droplet in the Leidenfrost state, and demonstrate in due course the occurrence of the sequential process of the initial nucleation of the oil microdroplets/nanodroplets; the self-encapsulation of the parent drop by an oil cap, which emerged out of the coalesced oil droplets, and finally, the micro-explosion of the parent drop.

\section*{Results and Discussion}
\subsection*{Four Life Stages of a Leidenfrost Ouzo Drop.}

\begin{figure*}[!htb]
	\centering
	\includegraphics[width=1 \linewidth]{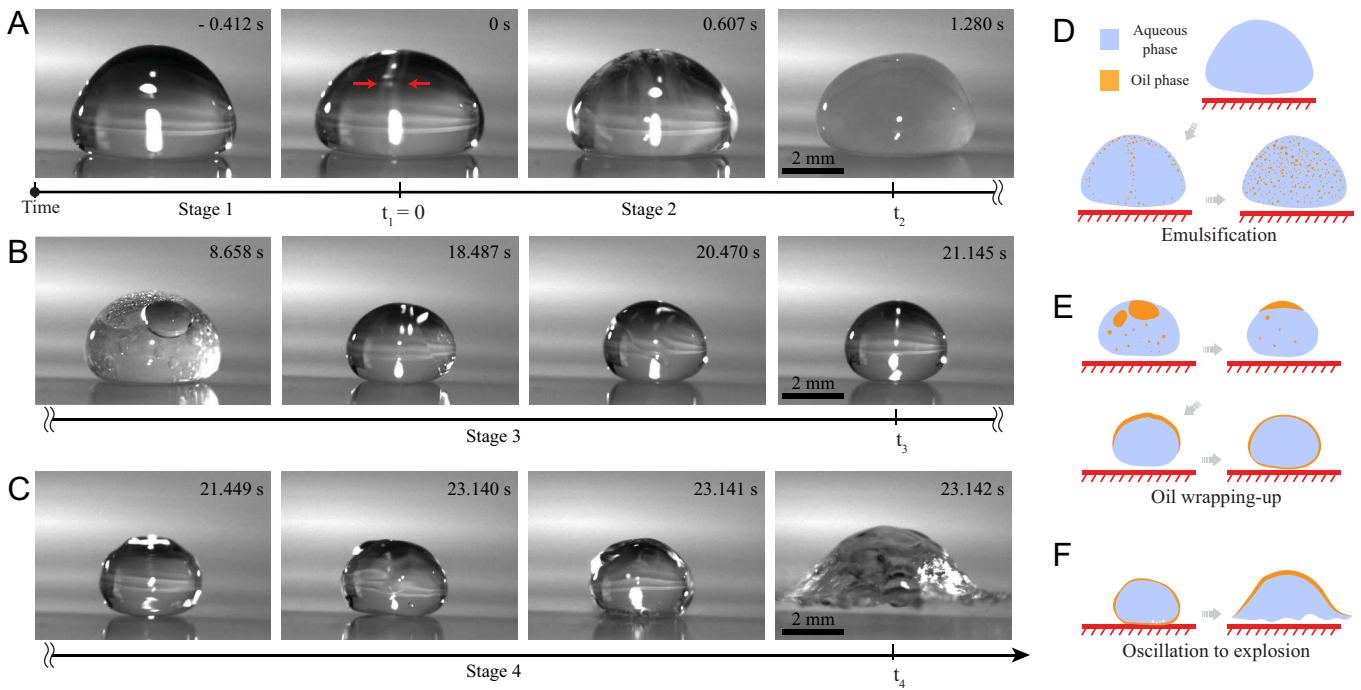}
	\caption{The entire boiling process of an ouzo drop initially in the Leidenfrost state can be divided into four stages. For each row, A-C are experimental results, and D-F are the corresponding sketches in the cross-section view. In the sketches, the orange parts represent the oil-rich phase and the blue parts the water-rich phase.
		$t_1, t_2, t_3$ and $t_4$ indicate the end moment of each stage. (A and D) Stages 1 and 2: The drop transforms from the transparent state to the early milky state. Oil microdroplets/nanodroplets first nucleate on the interface (see arrows), making the end moment $t_1$ of stage 1, defined as $t_1=0$. Then the drop is filled with oil emulsions and, accordingly, becomes fully opaque at $t_2$. (B and E) Stage 3: Tiny oil microdroplets gradually coalesce and, finally, form an oil cap on the upper half of the drop. The drop transforms from the early-milky state to a transparent state with two separate phases. The oil cap tries to cover the drop and, finally, wraps the drop at $t_3$. (C and F) Stage 4: The drop, which previously was in  a stable film-boiling Leidenfrost state, now becomes unstable. The vapor generated from the oil-encapsulation is no longer sufficient to levitate the drop, and consequently, the drop directly contacts the solid surface and explodes at $t_4$. (Scale bars: 2 mm.)}
	\label{fig1}
\end{figure*}

A millimeter drop of a miscible ouzo solution (with three components consisting of water, ethanol, and oil, see Materials and Methods for details) is deposited with a microliter-pipette onto a superheated surface at \SI{400}{\degreeCelsius}.
As expected, the ouzo drop is in the levitated Leidenfrost state (Fig.~\ref{fig1}A), and a vapor layer is formed below it, isolating it from direct contact with the solid surface and, thus, constituting the so-called film boiling state. A slightly curved quartz lens is used as the solid surface to prevent the drop from moving around, as detailed in the Method section.

The observed boiling process takes more than 20 seconds, with the evaporating ouzo drop experiences four distinct stages.
Fig.~\ref{fig1} shows representative experimental snapshots of the drop in each stage from a side view recording, together with the corresponding sketches in the cross-section view.
As shown in the first image of Fig.~\ref{fig1}A, the ouzo drop is initially miscible and hence optically transparent.
Subsequently, the preferential evaporation of the most volatile ethanol component reduces the oil solubility.
Consequently, spontaneous nucleation of the oil microdroplets/nanodroplets (arrows in the second image in Fig.~\ref{fig1}A)--- i.e., the ouzo effect---is triggered, as was similarly observed by Tan {\it et al}~\cite{tan2016evaporation,tan2017self} for an evaporating sessile ouzo droplet, {with the drop sitting on a solid substrate with a three-phase contact line.} This marks the end of the first stage. Many microdroplets appear and are advected around by the internal flow (the third image in Fig.~\ref{fig1}A). Then more and more oil microdroplets visibly appear in the drop---at the beginning only on the drop surface, but later, also in the bulk of the drop. After about 1.28 seconds, the drop is full of the oil microdroplets and becomes opaque (the fourth image in Fig.~\ref{fig1}A), defined as the end of stage 2. This is, again, similarly seen by Tan {\it et al}~\cite{tan2016evaporation,tan2017self} for an evaporating ouzo drop on a solid substrate. The entire process of the first two stages is sketched in Fig.~\ref{fig1}D. 
Starting from stage 3, the oil microdroplets start to coalesce and form large and transparent oil droplets, as shown in the first image of Fig.~\ref{fig1}B. These gradually merge into one entity, which constitutes part of the drop surface. Finally, an oil cap is formed on the upper half of the clear aqueous drop, as shown in the second image of Fig.~\ref{fig1}B. 
Remarkably, the oil cap then gradually spreads over the drop surface (the third image in Fig.~\ref{fig1}B), as also seen for a drop on an atmospheric surface~\cite{li2020evaporating,tan2019porous}. The process of the third stage is sketched in Fig.~\ref{fig1}E.
After the wrapping-up process is finished, a fully oil-capsulated Leidenfrost drop has been established. However, this aqueous-in-oil drop cannot be stably levitated, due to the lack of sufficient vapor, and leads to the last stage of the drop's life, as shown in Fig.~\ref{fig1}C. 
Specifically, in the first image in Fig.~\ref{fig1}C, the top part of the drop appears blurry, due to the unstable motion of the drop surface. With the levitating vapor layer formed by the vaporizing volatile water being cut off by the encapsulating surface oil layer, the droplet falls down and contacts the superheated surface. The droplet is now heated rapidly, eventually leading to the nucleation of the encapsulated volatile liquid core and, hence, {the} explosion of the drop as the end of its life, as shown with the third and fourth images of Fig.~\ref{fig1}C and the sketches in Fig.~\ref{fig1}F.

\begin{figure}[!h]
	\centering
	\includegraphics[width=0.9 \linewidth]{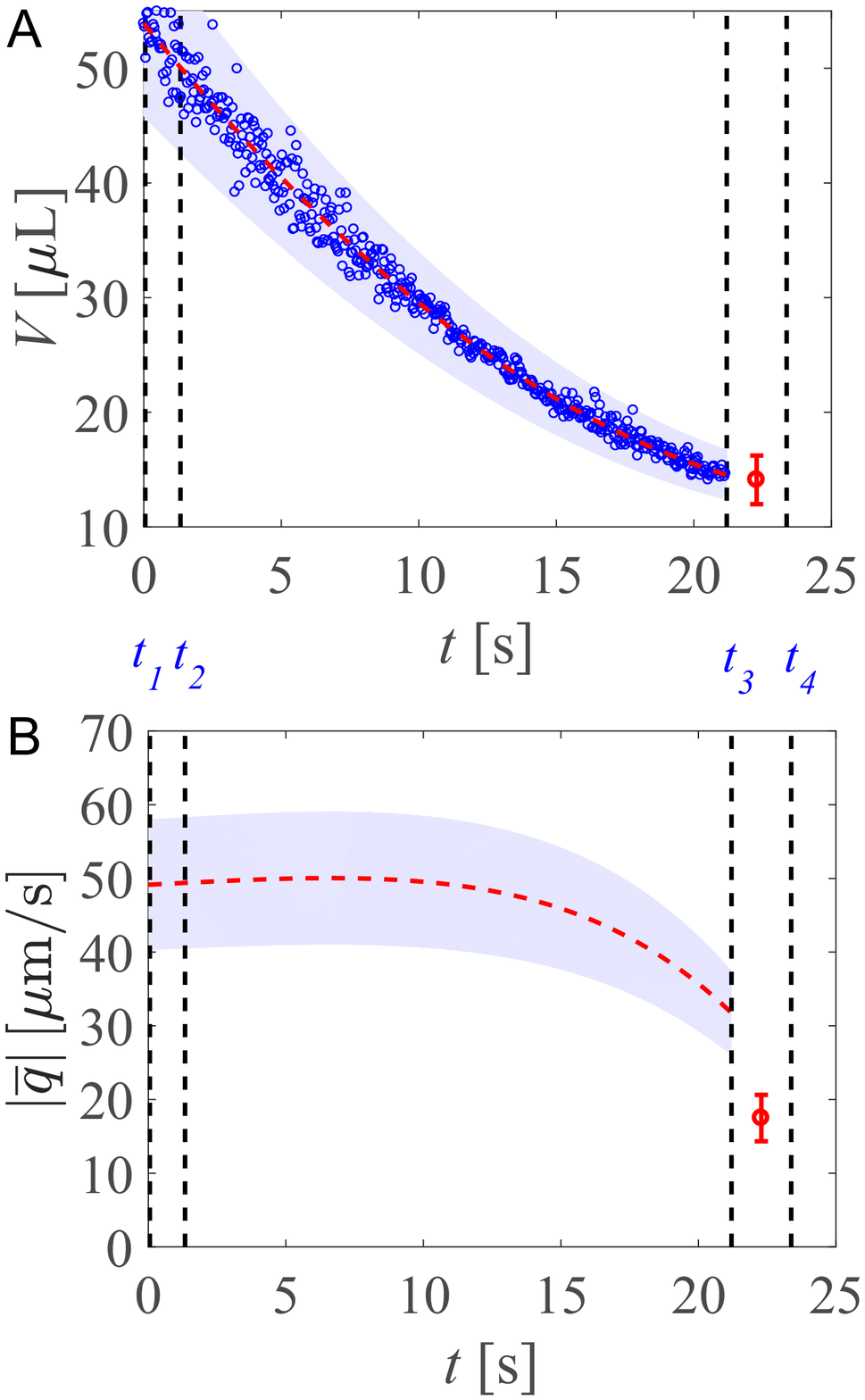}
	\caption{Quantitative analysis of the entire evaporation process. (A) Volume $V$ versus time $t$ of the ouzo drop. Blue circles show experimental results. The spreading of the data originates from the drop deformation, leading to an error in identifying the droplet volume from the cross-section (shown in the blue area). The red dashed line is the fitting line of the volume. The red circle with the error bar shows the average volume in stage 4. (B) Evolution of the absolute value of the evaporation volume flux rate per unit area $\left \vert{\overline{q} }\right \vert$ versus time. The red dashed line is calculated by the fitting lines of volume $V$ and entire area $S_a$ of the drop. The blue area shows the possible range of $\left \vert{\overline{q} }\right \vert$. The red circle with the error bar shows the average $\left \vert{\overline{q} }\right \vert$ in stage 4. Moments of $t_1$, $t_2$, $t_3$, and $t_4$ are marked through the vertical dashed lines. }\label{fig2}
\end{figure}

To start a quantitative analysis, we calculate the volume $V$ for the evolution of the Leidenfrost ouzo drop,
assuming that the drop maintains its axisymmetric shape during the entire evaporation process. As shown in Fig.~\ref{fig2}A, the drop size decreases rapidly within 24 seconds. Blue circles show experimental results. The scatter of the data originates from the drop deformation. The red dashed line and the blue area respectively show the fitting line and the changing range of the volume. It is seen that during stage 4, the drop is unsteady and asymmetric. Thus, it is hard to determine the exact volume evolution in time and only the drop volume before the explosion moment is calculated. The red circle with the error bar shows the average volume in stage 4. 
To gain an overall view of the evaporation rate, 
we calculate the absolute value of the volume decreasing rate per unit area, $\left \vert{\overline{q}}\right \vert =  \left \vert{{\mathrm{d}V}/{\mathrm{d}t}}\right \vert/S_a$, where $S_a$ is the entire surface area of the drop determined from the side profile of the drop.
As shown in Fig.~\ref{fig2}B, the red dashed line is calculated by the fitting lines of volume $V$ and the entire area $S_a$ of the drop. The blue area shows the possible range of $\left \vert{\overline{q} }\right \vert$. The red circle with the error bar shows the average $\left \vert{\overline{q} }\right \vert$ in stage 4. The range of the blue area and the value of the error bar are explained in the SI Appendix. It is seen that $\left \vert{\overline{q}}\right \vert$ initially does not change substantially, and then decreases rapidly. In stage 2, ethanol is the main evaporating component with a faster evaporation rate. In stage 3, water gradually becomes the main evaporating component with a slower evaporation rate than that of ethanol. As the oil film covers the water drop (between $t_3$, and $t_4$), the average evaporation rate decreases rapidly, as shown with the red circle in Fig.~\ref{fig2}B. The corresponding ending moments of each stage $t_1, t_2$, $t_3$, and $t_4$ are marked by the dashed vertical lines in Fig.~\ref{fig2}. {Three further and independent experimental realizations of the same experiment are shown in the SI Appendix, all giving similar results.}

As discussed above, the Leidenfrost ouzo drop experiences complex boiling dynamics, including four different stages, which we now discuss in more detail.

\subsection*{Stages 1 and 2: From Transparent Miscible Ouzo Drop to Emulsification due to Microdroplet Nucleation.}

\begin{figure*}[!htb]
	\centering
	\includegraphics[width=0.9 \linewidth]{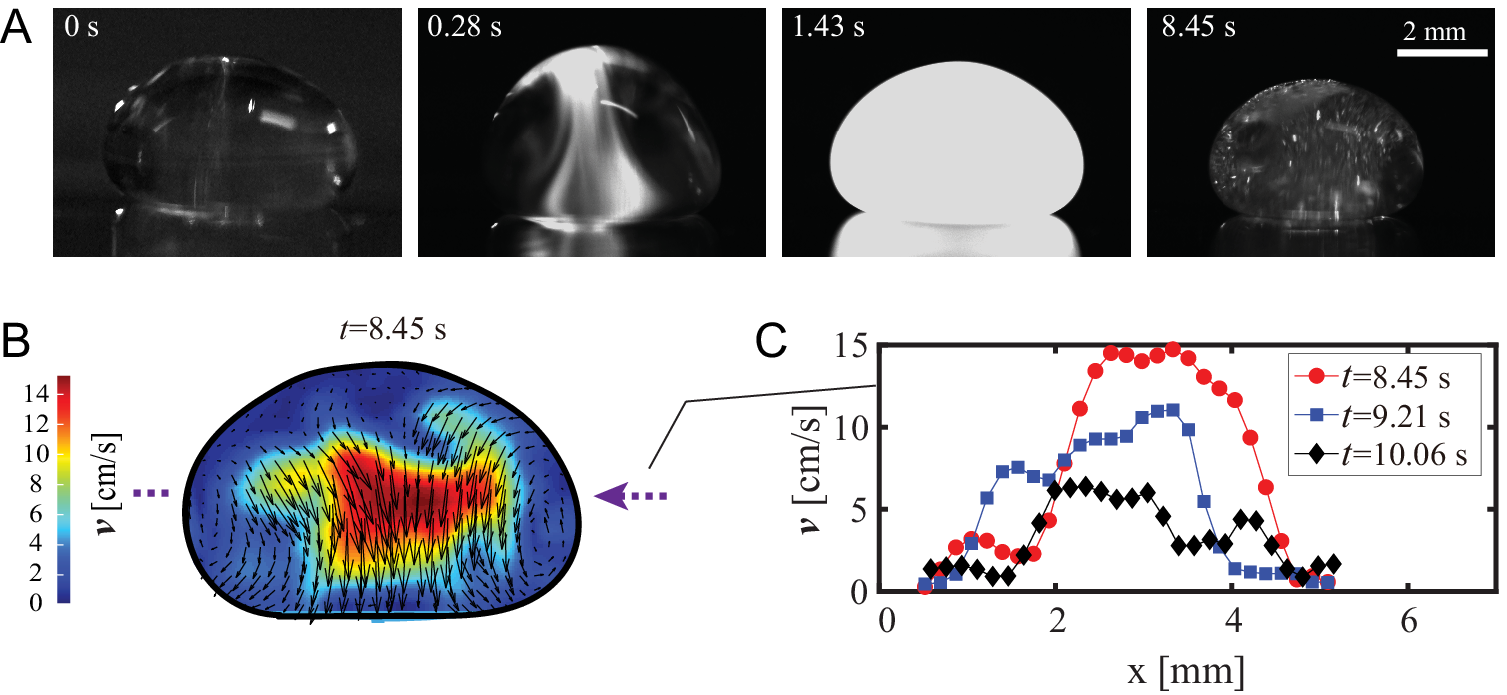}
	\caption{From transparent miscible ouzo drop to  emulsification due to microdroplet nucleation. (A) The white nucleated oil areas first occur on the interface and are advected by the internal flow. A laser sheet vertically shines through the center of the drop from the top for visualization. In the cross-section of the drop, the drop is transparent in the beginning. Then, the nucleated oil microdrops are distributed through two large vortexes. Subsequently, the oil microdroplets distribute throughout the entire drop. As these oil microdroplets coalesce, the drop becomes more transparent, and a strong flow is observed within the drop. (B, C) The images of A are used to analyze the velocity within the drop using PIV. (B) The internal velocity distribution in the cross-section of the drop shows that the flow moves downwards near the axis and upwards near the interface. Two large vortices form in the cross-section, corresponding to one toroidal vortex in the 3D drop. For all three times, the maximum velocity is around the center of the drop. At 8.45 s, the  maximum magnitude is about 15 cm/s. (C) {Temporal evolution} of the velocity at the half-height of the drop. With advancing time $t$, both the maximum velocity and mean velocity decrease. }
	\label{fig3}
\end{figure*}

The first characteristic event during the boiling process of the Leidenfrost ouzo drop is the transformation from the initial miscible drop to an opaque one due to the nucleated oil microdroplets, i.e., an emulsified drop.
The emulsification is due to the higher evaporation rates of ethanol, which leads to lower oil solubility and eventually, nucleation of oil microdroplets~\cite{tan2016evaporation}.

The temperature of the Leidenfrost drop can be approximated by the boiling temperature of the mixture~\cite{burton2012geometry}.
At the initial drop composition, the boiling temperature is around \SI{80}{\degreeCelsius}~\cite{noyes1901boiling}, which is higher than the boiling point of pure ethanol (\SI{78}{\degreeCelsius}), but lower than that of pure water (\SI{100}{\degreeCelsius}), and much lower than that of trans-anethole (\SI{234}{\degreeCelsius}).
Consequently, ethanol preferentially vaporizes, and its reduced concentration leads to the corresponding reduction of the oil solubility, which in turn induces the ouzo effect~\cite{tan2016evaporation} characterized by the spontaneous nucleation of oil emulsion.
Fig.~\ref{fig3}A gives four snapshots, showing the oil microdroplets start nucleating near the drop surface and then are advected by the convection inside the drop.

We use the microdroplets as tracer particles to reveal the flow field in the drop by vertically shining a laser sheet at the center of the drop.
A particle-image velocimetry (PIV) calculation (PIVlab) shown in Fig.~\ref{fig3}B displays the motion of the microdroplets, which move upwards along the surface and then downwards inside the drop {(Movie S1)}. Two large vortices form in the cross-section, corresponding to one toroidal vortex in the three-dimensional (3D) drop. 
For a gravity-flattened drop (see the third image in Fig.~\ref{fig3}A), the liquid at the drop base is entrained by the viscous flow in the vapor layer from the base center to the periphery, which contributes to the flow moving upwards near the interface and downwards near the axis~\cite{bouillant2018leidenfrost}. 
In addition, the temperature is lower at the apex of a large Leidenfrost drop~\cite{bouillant2018leidenfrost}, again giving a higher surface tension there. 
The temperature gradient also causes thermal Marangoni flows along with the interface from the base to the apex~\cite{bouillant2018leidenfrost}. 
However, a higher temperature at the drop base may induce a higher evaporation rate of ethanol, giving a higher surface tension there. 
{The concentration gradient may induce solutal Marangoni flows} along with the interface in the opposite direction of the thermal Marangoni flow. Apparently, the measured motion of the microdroplets (Fig.~\ref{fig3}B) suggests that solutal Marangoni flows are weaker in the current situation.

\subsection*{Stage 3: Water-in-Oil Drop by Oil Wrapping-Up.}

\begin{figure*}[!htb]
	\centering
	\includegraphics[width=0.7 \linewidth]{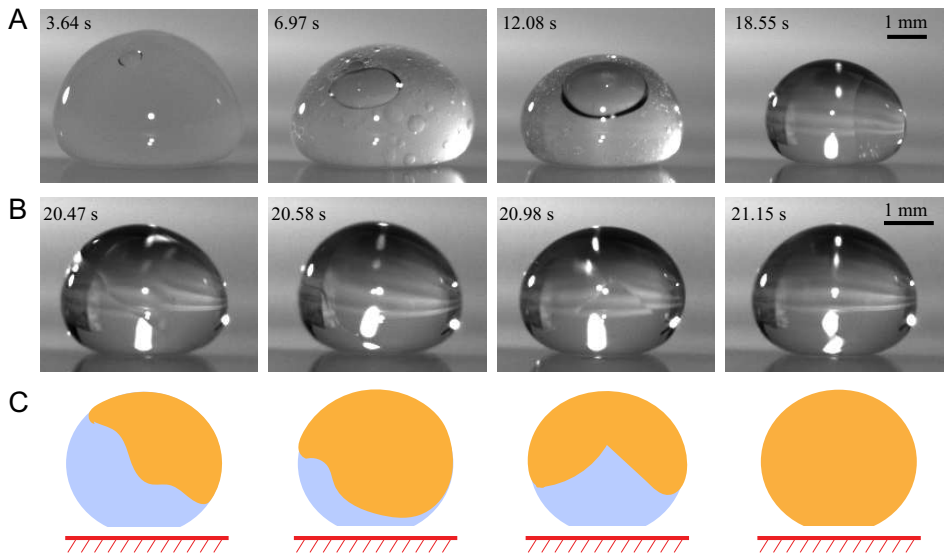}
	\caption{Stage 3 in the drop history. (A) The internal oil microdroplets coalesce, and correspondingly the mean size of oil droplets increases with time. The drop becomes more and more transparent. (B) The oil cap gradually covers the aqueous drop, and finally wraps around it. (C) The corresponding sketches of the wrapping process in the side view. Note that water is trapped inside the droplet. (Scale bars: 1 mm.) }\label{fig4}
\end{figure*}

\begin{figure*}[hbt!]
	\centering
	\includegraphics[width=1 \linewidth]{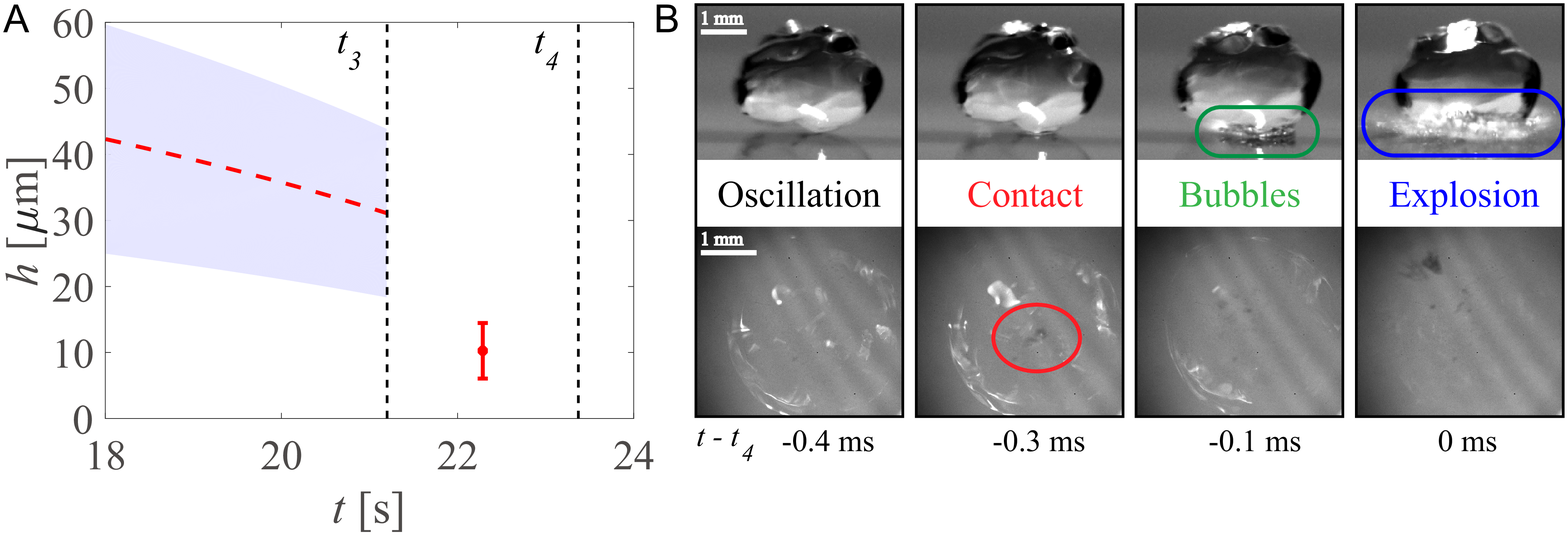}
	\caption{Stage 4 in the drop history. (A) Calculated vapor-layer thickness (based on Eq. 1) versus time during the end of stage 3 and in stage 4. The final vapor-layer thickness is above 30 $\mu$m in stage 3, but it decreases to around 10 $\mu$m in stage 4. The red dashed line and the blue area, respectively, show the average value and the changing range of the vapor-layer thickness. The red point with the error bar indicates the average vapor-layer thickness in stage 4. (B) Explosion of the drop. Combining the side (B, Upper) and the bottom (B, Lower) views, after the drop violently oscillates for a while, it locally contacts the heated substrate (shown in the red circle). Vapor bubbles are generated within the drop at the contact area (shown in the green circle). The vapor bubbles formed have large internal pressure, which causes a violent explosion in the end (shown in the blue circle). (Scale bars: 1 mm.) }\label{fig5}. 
\end{figure*}

The opaque emulsified drop is in a transient state, as the nucleated oil microdroplets are metastable.
The continuing oil nucleation and the microdroplet coalescence lead to the growth of the microdroplets.
However, the larger oil microdroplets cannot follow the flow closely.
As shown in Fig.~\ref{fig4}A, because of buoyancy they tend to float around the apex of the drop and consequently the oil phase merges at the top of the drop. 

To provide some insight into the merging and floating processes of the oil microdroplets, we consider the motion of the microdroplets that is controlled by two forces---i.e., the drag force $F_d \propto \mu R_o v_o$ and the buoyancy force $F_b \propto \Delta\rho R_o^3$, where $\mu$ is the liquid dynamic viscosity in the drop, and $R_o$ and $v_o$ are the radius and velocity of the microdroplets. Thus, the drag-buoyancy force ratio is inversely proportional to the droplet size squared, $F_d/F_b \propto \mu v_o \Delta\rho^{-1}R_o^{-2}$.
When the microdroplets are small, the drag force prevails, and hence, they follow the flow closely. 
However, the buoyancy force starts to dominate for the larger microdroplets. {Fig.~\ref{fig3}C shows the temporal evolution of the velocity at the half-height of the drop. For all three times, the maximum velocity is around the center of the drop. At 8.45~s, the maximum magnitude is about 15 cm/s. Recognizing that since most ethanol may be depleted at this time, to evaluate the dimensionless parameter of the water-rich aqueous drop, we can use the physical properties of liquid water at the boiling temperature. The P\'eclet number, representing the ratio of convective transfer over the diffusive one, is $Pe = v_{max}R/\alpha \approx 1741$, where $v_{max}$ is the maximum velocity within the drop, $R$ the equivalent radius of the drop, and $\alpha$ the thermal diffusivity of liquid water. Here we only consider the thermal transport, as the thermal effect is the dominant one in the current stage. The advective transport rate is much higher than that of thermal diffusion, suggesting that the diffusion is way too slow to level out temperature differences.} Furthermore, both the maximum velocity and mean velocity decrease with time. Since the small microdroplets grow and the average driven flow slows down, more and more oil droplets float up and accumulate at the top. 
As demonstrated in Fig.~\ref{fig4}A, the oil droplets float around the apex of the drop, and then gradually merge into an oil cap. 
{The phenomenon is new as compared to the evaporation of an ouzo drop directly in contact with a solid surface, as there, the contact line has major effects on the evaporation process~\cite{tan2016evaporation,tan2017self}. In such a case, the oil droplets accumulate at the contact line and form an oil ring. In contrast, here, the ternary Leidenfrost drop does not have a contact line, and the oil droplets can move freely within the Leidenfrost drop. Oil droplets start to coalesce and gradually merge into one entity.} Finally, the water-rich aqueous drop is attached to an oil cap.

The reduction of the ethanol concentration leads to an increase of the drop surface energy and correspondingly changes the wetting condition of the oil cap on the drop.
Recognizing that the surface tension of water is very high,  the oil cap tries to cover the aqueous drop to reduce the surface energy. 
Fig.~\ref{fig4}B then shows a series of snapshots that capture the spreading of the oil cap on the drop surface.
Additionally, the upward interfacial flow competes with the spreading of the oil from the top, which causes repeated advancing and retreating spreading of the oil film {(Movie S2)}.
In the end, the oil film spreads and wraps the entire drop. 
The sketches in Fig.~\ref{fig4}C show the wrapping process of the oil film as discussed. 
Consequently, the emulsified drop completes the oil-encapsulation and becomes a water-in-oil Leidenfrost drop. We note in passing that a similar wrapping-up phenomenon has also been observed in other fluids systems by Li {\it et al}~\cite{li2020evaporating}, Tan {\it et al}~\cite{tan2019porous}, and Kadota {\it et al}~\cite{kadota2007microexplosion}.

\subsection*{Stage 4: The Final Drop Explosion.}

According to the lubrication approximation, the lifting force induced by the vapor layer should balance the weight of the drop. The thickness of the vapor layer can then be evaluated as~\cite{lyu2019final}
\begin{equation}
h = \left( \dfrac{3\pi}{2}\dfrac{\mu_{v} \left \vert{\overline{q}}\right \vert \rho_{l} r_b^4}{\rho_{v}G } \right)^{1/3} \label{h},
\end{equation}
where $r_b$ is the horizontal extent of the drop bottom surface, and $\mu_{v}$, $\rho_{v}$, $\rho_{l}$ are the dynamic viscosity of vapor, the vapor density and the liquid-phase density of the evaporating component (water in stage 3 and oil in stage 4), respectively. The physical properties of water and oil are given in the SI Appendix. Because the volume of oil is much smaller than that of water and the density difference between oil and water is small, the gravity of the drop can be calculated by $G = \rho_{l,w}Vg$, where $\rho_{l,w}$ is the density of liquid water, $g$ is the gravitational acceleration.

As shown in Fig.~\ref{fig5}A, the vapor-layer thickness near the end of the last stage 3 is above 30  $\mu$m when water is the main evaporation phase. However, the thickness decreases to around 10 $\mu$m based on Eq.~\ref{h} when the drop is wrapped with the oil phase in stage 4, as shown in Fig.~\ref{fig5}A. The range of the blue area and the value of the error bar are explained in SI Appendix. The very thin vapor layer results from the extremely slow evaporation rate of the oil phase, as shown in Fig.~\ref{fig2}B. Such a thin vapor layer cannot stably levitate a millimeter-sized drop \cite{burton2012geometry}. A small perturbation will induce direct contact between the drop and the superheated solid surface, and this local contact consequently triggers microexplosion of the water droplet fully entrapped by the oil layer.

We combine the bottom and side recordings, as presented in Fig.~\ref{fig5}B, to identify the moment around the explosion.
In the first column, there is no contact yet, as shown in Lower. Then in the second column, the dark area marked in the red circle demonstrates that the drop contacts the heated substrate. Once the drop contacts the superheated substrate, the temperature of the contacting liquid increases rapidly above the oil boiling point, $T_o$, which is much higher than that of water, $T_w$. Thus, the water phase next to the oil contacting area is further superheated, and vapor bubbles are observed in the green circle of Fig.~\ref{fig5}B. 
Once the drop is unable to sustain the inner high pressure of the vapor bubbles \cite{prosperetti2017vapor}, the drop undergoes a violent explosion (see the blue circle of Fig.~\ref{fig5}B).

\section*{Conclusions and Outlook}

In conclusion, we have experimentally investigated the entire gasification process of a multicomponent Leidenfrost drop of ouzo composition. 
It is found that, upon gasification, the preferential evaporation and, hence, concentration reduction of the most volatile ethanol leads to the massive formation of oil microdroplets, which, in turn, agglomerate and form a surface layer that encapsulates the remaining aqueous drop. The aqueous drop, within the encapsulating oil layer, is then heated to the limit of superheat and consequently, nucleates near the superheated surface, hence rupturing the entire droplet. 

{The found phenomena illustrate the beauty \& richness of multicomponent drop systems with phase transition, which are also relevant in many combustion processes.}

\section*{Materials and Methods}
\subsection*{Experimental Setup and Procedures}
A high-speed camera (Photron Fastcam NOVA S12) with macro lens (Nikon 105 mm) was placed in the side view, and another high-speed camera (Photron Fastcam AX200) with a long-distance microscope (Navitar) was placed in the bottom view to simultaneously record the entire process. A slightly curved quartz lens was used as a substrate to limit the movement of drops. A sapphire base was placed under the quartz lens to improve the temperature uniformity of the substrate. Both of them were heated by an aluminum block heater. The surface temperature of the substrate, $T_s$, was measured by a thermocouple attached to the surface. The transparent ouzo drop solution was prepared with an initial composition of 31.80$\%$ (volume [vol]/vol) ultrapure water, 66.05$\%$ (vol/vol) ethanol ($\geq$99.7$\%$), and 2.15$\%$ (vol/vol) transanethole oil (Sigma-Aldrich; 99$\%$), which located the liquid to be initially in the one-phase regime and can easily enter the nucleation region~\cite{tan2017self,tan2016evaporation}. A large ouzo drop produced by a microlitre-pipette was deposited on the superheated substrate. {The initial volume of the drop was 100 $\mu$L. The quantitative calculation starts around 50 $\mu$L}, and the substrate temperature was around 400$^\circ$C.

\subsection*{Image analyses}
The images were analyzed using a Matlab code. The shape of the drop was assumed to be axisymmetric~\cite{ma2018self}, and the calculation on the volume evolution starts from stage 2.

\subsection*{ACKNOWLEDGMENTS}
We thank X. Chao, Y. C. Liu, and Y. S. Li for insightful discussions. The work was support by Natural Science Foundation of China Grant 11988102, National Natural Science Foundation of China Joint Research Program 11861131005, Deutsche Forschungsgemeinschaft Program OH 75/3-1, and Tsinghua University Initiative Scientific Research Program 20193080058.




\end{document}